\documentclass[fleqn,usenatbib]{mnras}

\usepackage{newtxtext,newtxmath}

\usepackage[T1]{fontenc}
\usepackage{ae,aecompl}
\usepackage{times}

\usepackage{graphicx}	
\usepackage{amsmath}	
\usepackage{amssymb}	

\usepackage{symbols}
\usepackage{enumitem}

\usepackage{etoolbox}
\makeatletter
\patchcmd\@combinedblfloats{\box\@outputbox}{\unvbox\@outputbox}{}{%
}%
\makeatother

\title[When star formation drives the baryon cycle]{The Elephant in the Bathtub: when the physics of star formation regulate the baryon cycle of galaxies}

\author[J. Gensior and J.\ M.\ D.\ Kruijssen]{
Jindra Gensior\thanks{E-mail: j.gensior@uni-heidelberg.de} and
J. M. Diederik Kruijssen
\\
Astronomisches Rechen-Institut, Zentrum f{\"u}r Astronomie der Universit{\"a}t Heidelberg, M{\"o}nchhofstra{\ss}e 12-14, 69120 Heidelberg, Germany
}

\date{Accepted 2020 November 03. Received 2020 November 02; in original form 2020 July 21}

\pubyear{2020}

\begin{document}
\label{firstpage}
\pagerange{\pageref{firstpage}--\pageref{lastpage}}
\maketitle

\begin{abstract}
In simple models of galaxy formation and evolution, star formation is solely regulated by the amount of gas present in the galaxy. However, it has recently been shown that star formation can be suppressed by galactic dynamics in galaxies that contain a dominant spheroidal component and a low gas fraction. This `dynamical suppression' is hypothesised to also contribute to quenching gas-rich galaxies at high redshift, but its impact on the galaxy population at large remains unclear. In this paper, we assess the importance of dynamical suppression in the context of gas regulator models of galaxy evolution through hydrodynamic simulations of isolated galaxies, with gas-to-stellar mass ratios of 0.01--0.20 and a range of galactic gravitational potentials from disc-dominated to spheroidal. Star formation is modelled using a dynamics-dependent efficiency per free-fall time, which depends on the virial parameter of the gas. We find that dynamical suppression becomes more effective at lower gas fractions and quantify its impact on the star formation rate as a function of gas fraction and stellar spheroid mass surface density. We combine the results of our simulations with observed scaling relations that describe the change of galaxy properties across cosmic time, and determine the galaxy mass and redshift range where dynamical suppression may affect the baryon cycle. We predict that the physics of star formation can limit and regulate the baryon cycle at low redshifts ($z \lesssim 1.4$) and high galaxy masses ($\Mstar \gtrsim 3 \times 10^{10}~\Msun$), where dynamical suppression can drive galaxies off the star formation main sequence. 
\end{abstract}

\begin{keywords}
galaxies: elliptical and lenticular, cD -- galaxies: evolution -- galaxies: ISM -- galaxies: star formation 
\end{keywords}



\section{Introduction}
The past 10-15 years have brought major progress in understanding the scaling relations describing the galaxy population. Star-forming galaxies are organised along the star formation main sequence (SFMS; see e.g.\ \citealt{Noeske2007,Peng2010}), which is a tight relation between galaxy mass and star formation rate (SFR). The main source of progress has been the development of simple models describing the baryon cycle of galaxies as they evolve over cosmic time. These models relate the galactic-scale processes of star formation and stellar feedback to gas accretion onto dark matter haloes. A family of galaxy evolution models referred to as `bathtub' or `gas-regulator' models \citep[e.g.][]{Finlator2008,Bouche2010,Lilly2013,Dekel2013} do this by parametrising these complex physical processes in simple, analytically-solvable equations. In these models, the gas inflow onto a galaxy determines its properties and evolution. Star formation is solely dependent on the presence of cold gas that is converted into stars with a constant efficiency, as observed in most star-forming galaxies \citep[e.g.][]{Kennicutt1998,Leroy2013}. The star formation rate is thus regulated by the balance between gas inflow and the stellar feedback-driven outflow \citep{Hopkins2011,Agertz2013}.

Gas-regulator models are particularly compelling, because they allow one to interpret observational results and make predictions for the evolution of the galaxy population without requiring sophisticated semi-analytic models or full hydrodynamic simulations. For instance, they have been highly successful in reproducing and explaining scaling relations relating to the chemical enrichment of galaxies, like the mass-metallicity relation \citep{Lequeux1979,Tremonti2004} or the mass-metallicity-SFR relation \citep{Mannucci2010}. By contrast, the specific SFR (sSFR, the SFR normalised by the stellar mass of the galaxy) is reproduced well by regulator models at redshift $z \geq 3$, but discrepancies arise at lower redshifts \citep{Dekel2014}. Furthermore, gas-regulator models are focused on the star-forming population of galaxies, and it is not clear how they extend to the quiescent population of galaxies that are (in the process of being) quenched. \citet{Peng2014} suggest that this population could be accounted for by simply lowering the value of the star formation efficiency, which is otherwise assumed to be constant.

In the gas-regulator framework, the quenching of galaxies is a direct result of no longer hosting a reservoir of cold, molecular gas. This can be achieved through heating by active galactic nuclei (AGN) or by expelling the available gas \citep[see e.g.][]{Croton2006,Bower2006,Fabian2012,Lacerda2020}. Gas supplies can also be depleted through strangulation \citep{Larson1980} or halo quenching \citep{Dekel2006}. In galaxy clusters or groups, galaxies can be quenched by external effects removing the gas, such as ram pressure stripping \citep{Gunn1972} or harassment \citep{Moore1996}. However, it remains a major open question how galaxies quench and remain quiescent while retaining (a substantial amount of) molecular gas \citep[e.g.][]{Davis2014}, especially in isolated galaxies that do not experience strong AGN feedback. This cannot be accomplished through halo quenching on its own, because in that case galaxies should remain star-forming until all their molecular gas is depleted, contrary to the observations. \citet{Martig2009} proposed that a galaxy's gaseous component may be stabilised against gravitational collapse and star formation by the presence of a dominant spheroidal component. This process was originally dubbed `morphological quenching' and has been put forward as a possible quenching mechanism in spheroid-dominated galaxies. In this picture, the spheroid's gravitational potential increases the shear experienced by the gas, which in turn enhances its velocity dispersion and its \citet{Toomre1964} $Q$ parameter. This prevents the fragmentation of the gas into dense clouds and subsequently suppresses star formation.  

Isolated, early-type galaxies in the local Universe exhibit SFRs per unit molecular gas mass that are suppressed compared to late-type galaxies \citep[compare][]{Bigiel2008,Davis2014}. It has been proposed that morphological quenching is responsible for the inefficient star formation in early-type galaxies \citep{Martig2013,Davis2014}. Morphological quenching has also been invoked as a quenching mechanism at intermediate redshifts of $0.6 \lesssim z \lesssim 1.2$ for galaxies with compact central bulges \citep{Kim2018}, and to explain the existence of two gas rich post-starburst galaxies at $z\sim 0.7$ \citep{Suess2017}. \citet{Genzel2014} showed that the Toomre $Q$ parameter increases towards the centres of galaxies at z$\sim$2 and interpreted this as an early sign of morphological quenching. Similarly, a number of studies have related the SFRs of galaxies at $z\sim 2{-}3$ to their morphologies and hypothesise that morphological quenching is at least partially responsible for the quiescent galaxy population \citep[e.g.][]{Wuyts2011,Barro2013,HuertasCompany2015,Barro2017}. 

In \citet{Gensior2020}, we set out to rigorously explore the phenomenological concept of morphological quenching\footnote{We will exclusively use the term dynamical suppression from here onward, as the galactic dynamics are a crucial component for the suppression and quenching, while morphological quenching places all emphasis on the presence of a bulge.} by running a suite of isolated galaxy simulations with a sub-grid star formation model that relates the star formation efficiency to galactic dynamics via a dependence on the virial parameter of the gas. The parameter space spanned by these simulations ranges from disc galaxies to spheroids and was designed to assess whether a dominant bulge component would be able to induce turbulence and suppress star formation in the galaxy, for Milky Way-mass galaxies with a gas to stellar mass ratio of $M_{\rm gas}/M_*=0.05$. In particular, we showed that the shear induced by the deep potential well of the bulge can drive up the gas velocity dispersion sufficiently to render the gas supervirial, thereby suppressing star formation in the central region. This tested and confirmed the hypothesis of \citet{Martig2009} that the median velocity dispersion of the gas increases towards the galactic centre in the presence of a bulge \citep{Gensior2020}. Both the increase in velocity dispersion and the subsequent suppression of star formation are enhanced at higher stellar surface densities ($\mus$), which is a proxy for the dominance of the bulge in the galactic potential. This is consistent with the dearth of star formation observed in a post-starburst galaxy with a gas fraction of 5~per~cent by \citet{Suess2017} and also agrees with the result of \citet{Kretschmer2020}, who found that dynamical suppression is sufficient to quench a galaxy in a cosmological zoom-in simulation following a merger-driven starburst and subsequent gas expulsion.

In the context of the aforementioned gas-regulator models, these results raise a critical question: is there any part of the galaxy population for which the physics of star formation can dominate the baryon cycle through dynamical suppression? If the answer is affirmative, this would necessitate the extension of gas-regulator models with an environmentally-dependent star formation efficiency, such that the SFRs of galaxies are not exclusively set by their gas accretion rates.

There are suggestions in the literature that dynamical suppression is sensitive to the gas fraction of the galaxy and ineffective if the galaxy has a large, or continuously replenishing gas reservoir. For instance, \citet{Martig2013} only report a suppression of star formation for their galaxy simulated with a gas fraction of 1.3~per~cent, but not for their galaxy with a gas fraction of 4.5~per~cent. However, their simulations use a constant star formation efficiency, which leads to galaxies with approximately the same SFR even if $\mus$ differs by $\sim 0.7$~dex. With this insensitivity to the galactic potential, simulations with a constant star formation efficiency consequently do not accurately reproduce the observed star formation suppression of spheroid-dominated galaxies \citep{Gensior2020}. None the less, it is plausible that the dynamical suppression of star formation is only triggered below some threshold gas fraction, where the gas no longer dominates the local gravitational potential. Indeed, simulations with more sophisticated sub-grid star formation models yield similar behaviour. Simulations of reionization with dynamics-dependent star formation indicate that at extremely high redshift ($z > 5.7$), with gas fractions exceeding 50~per~cent, the time-averaged star formation rate remains unaffected by the build-up of a bulge component \citep{Trebitsch2017}. Likewise, \citet{Su2019} performed simulations of isolated galaxies and found that continued accretion and cooling of gas from the hot circumgalactic medium prevents the galaxies from quenching, even if the presence of a bulge causes a mild suppression of star formation. 

In this paper, we investigate the impact of dynamical suppression on the galaxy population across cosmic time. We wish to establish whether there exist any conditions under which the physics of star formation represent the rate-limiting step in (and therefore regulate) the baryon cycle and, if so, determine the galaxy mass and redshift range for which this is predicted to occur. Therefore, we extend the parameter space covered in \citet{Gensior2020} by repeating our simulations of galaxies with different gravitational potentials for a wide variety of different gas fractions. In Section~\ref{s:Sims}, we outline the numerical methods used and introduce the simulations. The results are presented in Section~\ref{s:Res}. In Section~\ref{s:EiB}, we assess the impact of our results on the galaxy population by predicting at which galaxy masses and redshifts dynamical suppression should affect the SFR in galaxies. Finally, we summarise and discuss these results in Section~\ref{s:SD}.

\section{Simulations} \label{s:Sims}
All simulations analysed in this work were performed with the moving-mesh code {\sc{Arepo}} \citep{Springel2010}. In {\sc{Arepo}}, stars and dark matter are treated as Langrangian particles, while the equations of hydrodynamics are solved on an unstructured mesh, created from a Voronoi tesselation, with a second-order, unsplit Godunov solver. Gravitational interactions are calculated using a tree-based scheme.

Treatment of star formation, cooling and feedback is the same as in \citet{Gensior2020}, which we briefly summarise here. The sub-grid model for star formation uses a virial parameter ($\avir$) based star formation efficiency per free-fall time ($\eff$) as in \citet{Padoan2012,Padoan2017}:
\begin{equation} \label{eq:eff}
    \eff = 0.4 \exp{ \left(-1.6 \avir^{0.5}\right)} , 
\end{equation}
which we use to calculate the star formation rate density $\dot{\rho}_{\rm SFR}$ of a gas cell as
\begin{equation} \label{eq:SFR}
    \dot{\rho}_{\rm SFR} = \eff \frac{\rho}{\tff} .
\end{equation}
Equation~(\ref{eq:eff}) introduces a dependence on the gas dynamics into the \citet{Katz1992} star formation prescription of equation~(\ref{eq:SFR}), which otherwise just depends on the gas density $\rho$, both explicitly and through the free-fall time $\tff = \sqrt{3\pi/32G\rho}$. 
The virial parameter is calculated self-consistently using the density gradient based approach introduced by \citet{Gensior2020}, where
\begin{equation}
    \avir^{1/2} \propto \left|\frac{\dg}{\rho}\right|\frac{\sigma}{\rho^{1/2}},
\end{equation}\label{eq:avir}
which is calculated exclusively using the gas properties. This approach effectively performs an on-the-fly cloud identification by iterating over neighbouring gas cells until an overdensity is identified, from which the velocity dispersion $\sigma$ and subsequently the virial parameter are calculated. For consistency with the simulations described in \citet{Gensior2020}, we impose a minimum density threshold of $1~\ccm$ and a maximum temperature threshold of $1 \times 10^3~\K$ to define gas that is eligible for star formation. To model the thermal state of the gas, we use the Grackle chemistry and cooling library\footnote{https://grackle.readthedocs.io/} \citep{Smith2017}, with the six species chemistry network and tabulated atomic fine structure cooling at solar metallicity. A constant \citet{Haardt2012} UV-background is used to model heating due to the interstellar radiation field. Feedback from star formation is included as mechanical feedback from Type II supernovae (SNe) \citep{Kimm2014,Hopkins2014}. Each SN injects $1\times10^{51}~\ergs$ of energy and ejects $10.5~\Msun$ into the surrounding gas, of which $2~\Msun$ are metals. Numerically, this is done by using a kernel to deposit the feedback to the 32 nearest neighbours, similar to \citet{Hopkins2014}. We assume one SN per $100~\Msun$ stellar mass formed \citep{Chabrier2003, Leitherer2014} and a delay time of $4~\Myr$, in line with recently observed feedback disruption times \citep{Kruijssen2019,Chevance2020c,Chevance2020}, before detonation. 

\begin{table}
 \begin{tabular}{lcccc}
  \hline
  Name & M$_{\rm b}$ [10$^{10}\Msun$] & R$_{\rm b}$ [kpc] & f$_{\rm gas}$ \\
  \hline
  B\_M90\_R1\_fg1 & 4.24 & 1 & 0.01 \\
  B\_M90\_R1\_fg3 & 4.24 & 1 & 0.03 \\
  B\_M90\_R1\_fg5 & 4.24 & 1 & 0.05 \\
  B\_M90\_R1\_fg10 & 4.24 & 1 & 0.10 \\
  B\_M90\_R1\_fg20 & 4.24 & 1 & 0.20 \\
  B\_M60\_R2\_fg1 & 2.83 & 2 & 0.01 \\
  B\_M60\_R2\_fg3 & 2.83 & 2 & 0.03 \\
  B\_M60\_R2\_fg5 & 2.83 & 2 & 0.05 \\
  B\_M60\_R2\_fg10 & 2.83 & 2 & 0.10 \\
  B\_M60\_R2\_fg20 & 2.83 & 2 & 0.20 \\
  B\_M30\_R3\_fg1 & 1.41 & 3 & 0.01 \\  
  B\_M30\_R3\_fg3 & 1.41 & 3 & 0.03 \\  
  B\_M30\_R3\_fg5 & 1.41 & 3 & 0.05 \\ 
  B\_M30\_R3\_fg10 & 1.41 & 3 & 0.10 \\ 
  B\_M30\_R3\_fg20 & 1.41 & 3 & 0.20 \\
  \hline
  \hline
 \end{tabular}
 \caption{Initial conditions of the simulations. Simulations are named by their bulge component and gas fraction, i.e.\ prefix `B', followed by the relative bulge mass (`M$X$' with $X$ the percentage of the total mass constituted by the bulge), the bulge scale radius (`R$Y$' with $Y$ the radius in kpc) and lastly the ratio of gas to stellar mass (`fg$Z$' with $Z$ the ratio multiplied by 100).}
 \label{tab:ICs}
\end{table}

The initial conditions resemble those of the {\sc{Agora}} disc \citep{Kim2016}, and were chosen from the parameter space surveyed in \citet{Gensior2020} (specifically, we adopt and modify runs B\_M30\_R3, B\_M60\_R2 and B\_M90\_R1 from that work), to cover a range in different galactic potentials from disc-dominated galaxy to spheroidal. Table~\ref{tab:ICs} lists all isolated galaxy simulations considered in this paper.  
All initial conditions were created according to the procedure detailed by \citet{Springel2005}. Each isolated galaxy is initialised in a dark matter halo with a \citet{Hernquist1990} profile which has a concentration $c=10$, a spin parameter $\lambda = 0.04$ and a circular velocity $v_{\rm circ} = 180~\kms$. The stellar component has a mass of $\Mstar \sim 4.71 \times 10^{10}~\Msun$ and is divided into a bulge and disc component. We also model the bulge with a \citet{Hernquist1990} profile, while the disc is described by an exponential radial profile and a sech$^2$ profile in the vertical direction. The defining parameters of the bulge profile are varied between different initial conditions, from $30{-}90$~per~cent of the initial stellar mass for the bulge mass $\Mb$ and $1{-}3~\kpc$ for the bulge scale radius $\Rb$. To maximise the difference in gravitational potential between the three different options, we combine $\Mb$ and $\Rb$ to yield a compact, completely bulge-dominated galaxy ($\Mb=0.9\Mstar$, $\Rb=1~\kpc$), a disc-dominated galaxy ($\Mb=0.3\Mstar$, $\Rb=3~\kpc$) and one with an intermediate bulge component ($\Mb=0.6\Mstar$, $\Rb=2~\kpc$). The mass of stars in the disc is determined by $\Md = \Mstar - \Mb$, its radial scale length is $\Rd = 4.6~\kpc$, and its exponential scale height is $0.1\Rd$. The gaseous component follows the same profile as the stellar disc. We vary the mass of gas in the disc for each set of initial conditions between $0.01{-}0.20\Mstar$. This is done to explore how the trends with gravitational potential seen for $\fg\equiv M_{\rm gas}/\Mstar=0.05$ reported by \citet{Gensior2020} depend on the gas fraction, extending the suite into regimes where galaxies are more gas-poor or more gas-rich than the original sample (thereby smoothly connecting the properties of local early-type galaxies to those of high redshift galaxies, e.g.\ \citealt{Geach2011, Tacconi2013}). A mass resolution of $\sim 1 \times 10^4~\Msun$ is used for the gas cells and stellar particles, while that of the dark matter is $1 \times 10^5~\Msun$. We use adaptive gravitational softening \citep{Price2007} to achieve optimum gravitational resolution, with minimum softening lengths of 12 and 26 pc for baryons and dark matter, respectively. This results in an average softening length of 25 pc for gas at a density of $10~\ccm$.

\section{Results}\label{s:Res}
\begin{figure*}
    \centering
    \includegraphics[width=1.\linewidth]{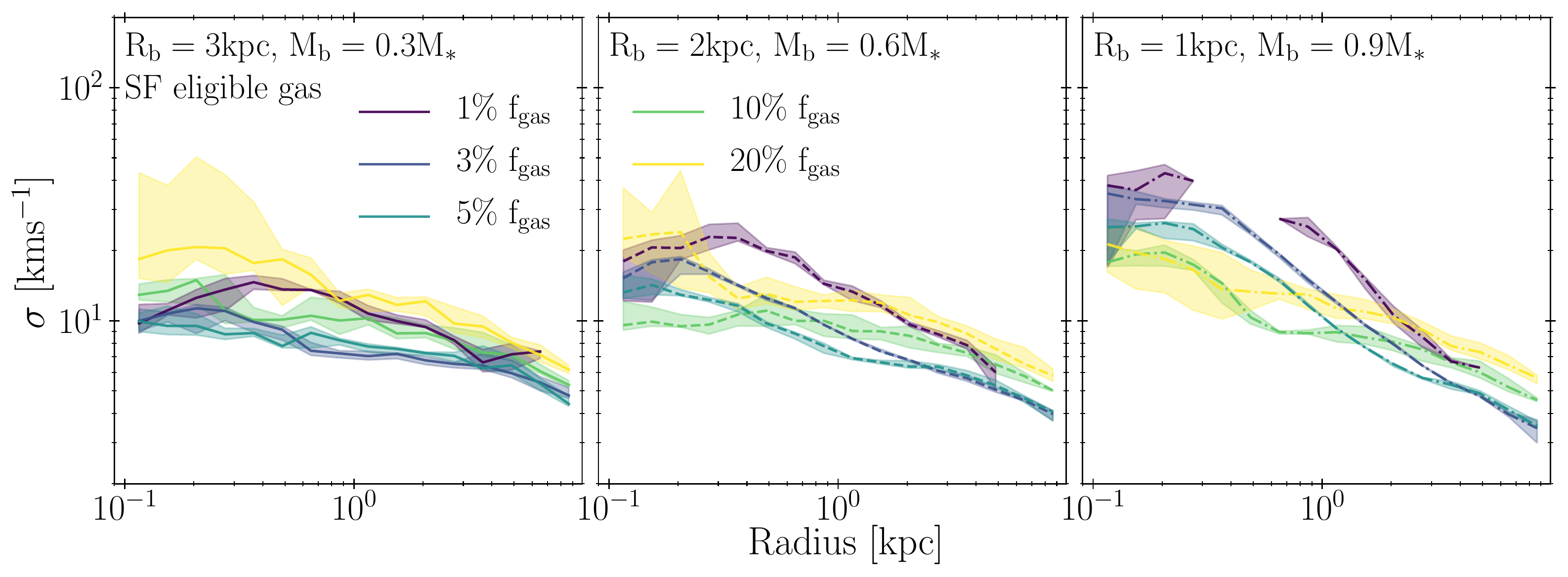}%
    \caption{The effect of the gravitational potential and gas fractions on the velocity dispersion. Each panel shows the radial profile of the gas velocity dispersion for galaxies with gas-to-stellar mass ratios of $0.01{-}0.20$, for the disc-dominated (left), intermediate bulge (middle) and completely bulge-dominated (right) potentials. Lines indicate the median over time, while the shaded regions indicate the 16th-to-84th percentile variation of the median over time.}
    \label{fig:veldisp_radprof}
\end{figure*}
\subsection{Gas dynamics}\label{ss:GD}
We first explore how the turbulent state of the gas is affected by the underlying gravitational potential and the gas fraction. The turbulent velocity dispersion directly influences the star formation rate via the virial parameter. Figure~\ref{fig:veldisp_radprof} shows the radial velocity dispersion profiles for gas meeting our density and temperature criteria for star formation. We measure the median and the 16th-to-84th percentiles in several snapshots to average over fluctuations due to cloud-scale evolutionary cycling \citep[e.g.][]{Kruijssen2018,Chevance2020b}. To do so, we use snapshots from $300~\Myr$ onward (offset from the start of the simulation to allow the galaxy to settle into equilibrium) until the end of the simulation at $1~\Gyr$. The snapshots are separated by $100~\Myr$, which is approximately a galactic dynamical time.

The panels in Figure~\ref{fig:veldisp_radprof} represent different gravitational potentials, with disc-dominated potential (runs B\_M30\_R3) on the left, intermediate bulges (runs B\_M60\_R2) in the middle, and the extremely bulge-dominated potential (runs B\_M90\_R1) on the right. In all simulations, the velocity dispersion increases towards the galactic centre. For all gas-to-stellar mass ratios $\fg\leq0.10$, this increase is steeper for more prominent bulges. The runs with $\fg=0.20$ are the only exception, because their median velocity dispersion profiles are the same across all panels to within the snapshot-to-snapshot time variation. We thus identify a gas-to-stellar mass ratio threshold of $\fg \sim 10$~per~cent) above which the velocity dispersion of the gas disc is no longer affected by the underlying gravitational potential. Even in the run with the most compact bulge (B\_M90\_R1\_fg20), the gas disc is sufficiently dense that its self-gravity dominates over the shear induced by the stellar potential.

All runs with a gas-to-stellar mass ratio of $\fg=0.01$ exhibit a similar velocity dispersion profile shape. The median velocity dispersion increases from the outskirts of the disc towards the centre of the galaxy\footnote{The profile of B\_M90\_R1\_fg1 contains a gap between $300{-}700~\pc$. This is a result of the star formation eligibility criteria that we apply. Although gas is present at all radii within the simulation, the shear induced by the potential prevents gas from fragmenting into dense clumps and instead smooths it into a featureless disc \citep{Gensior2020}. At $\fg=0.01$, this leads to the absence of star-formation eligible gas between $300{-}700~\pc$, because the shearing apart of overdensities leads to gas with densities below $1~\ccm$ when so little gas is present. For a quantitative discussion of how the these mechanisms affect the gas density distribution and the interstellar medium (ISM) structure in the simulations, we refer the interested reader to Appendix~\ref{A:gas}.}. As we demonstrate in \citet{Gensior2020} this radial trend signifies that the shearing motions in the underlying potential increase the velocity dispersion. It demonstrates that at the lowest gas fractions a minor bulge component is enough to affect the gas dynamics. Even for the disc-dominated potential (B\_M30\_R3\_fg1), we find a modest increase in velocity dispersion, of $\sim 0.3$ dex across the full range of radii, whereas the bulge-dominated potential (B\_M90\_R1\_fg1) induces a $\sim1$~dex difference, sustaining median velocity dispersions of $\sim 40{-}50~\kms$ at the galactic centre.

At intermediate gas-to-stellar mass ratios ($\fg=0.03{-}0.10$), the gas in the disc-dominated runs (B\_M30\_R3, left panel in Figure~\ref{fig:veldisp_radprof}) has an approximately flat median velocity dispersion throughout the inner $\sim 5\kpc$, which only declines slightly towards larger radii. This is equivalent to the profile found for a bulgeless galaxy by \citet{Gensior2020}. Conversely, the most bulge-dominated simulations (B\_M90\_R1, right panel in Figure~\ref{fig:veldisp_radprof}) all show the distinct increase in velocity dispersion towards the centre of the galaxy indicative of turbulent motions induced by the bulge. Also at these intermediate gas fractions, we find that the magnitude of the effect increases towards lower gas fractions. The velocity dispersion increases by $\sim \{0.9,0.7\}$~dex within the inner $\{4, 3\}~\kpc$ in the runs with $\fg=\{0.03, 0.05\}$, while the run with $\fg=0.10$ experiences a rise of 0.3~dex in the inner 700~pc. Under the influence of the intermediately bulge-dominated potential (B\_M60\_R2, middle panel in Figure~\ref{fig:veldisp_radprof}), only runs with $\fg \leq 0.05$ show an increase of the central velocity dispersion. This shows that the gas-to-stellar mass threshold below which the bulge enhances the gas velocity dispersion increases with the central stellar surface density. 

At high gas-to-stellar mass ratios ($\fg=0.10{-}0.20$), the median velocity dispersion profiles are nearly insensitive to the gravitational potential. Interestingly, they also exhibit a larger time variation of the median velocity dispersion compared to those with $\fg \leq 0.05$ for all gravitational potentials. This is related to the significantly higher SFR in these objects (also see Section~\ref{ss:SF}). One might expect that a larger global SFR leads to less bursty star formation and thus less variation of the median velocity dispersion over time. However, within each radial bin, star formation is always bursty and thus a net high(er) SFR leads to a larger number of subsequent feedback events and more variation of the velocity dispersion. An additional difference compared to the profiles of the galaxies with $\fg\leq0.05$ is that they have slightly elevated velocity dispersions at large radii. A higher gas fraction leads to a more massive gas disc and will increase the mid-plane pressure, against which the gas needs to support itself with higher $\sigma$ \citep[e.g.][]{Krumholz2018,Schruba2019}. 

\begin{figure*}
    \centering
    \includegraphics[width=0.47\linewidth]{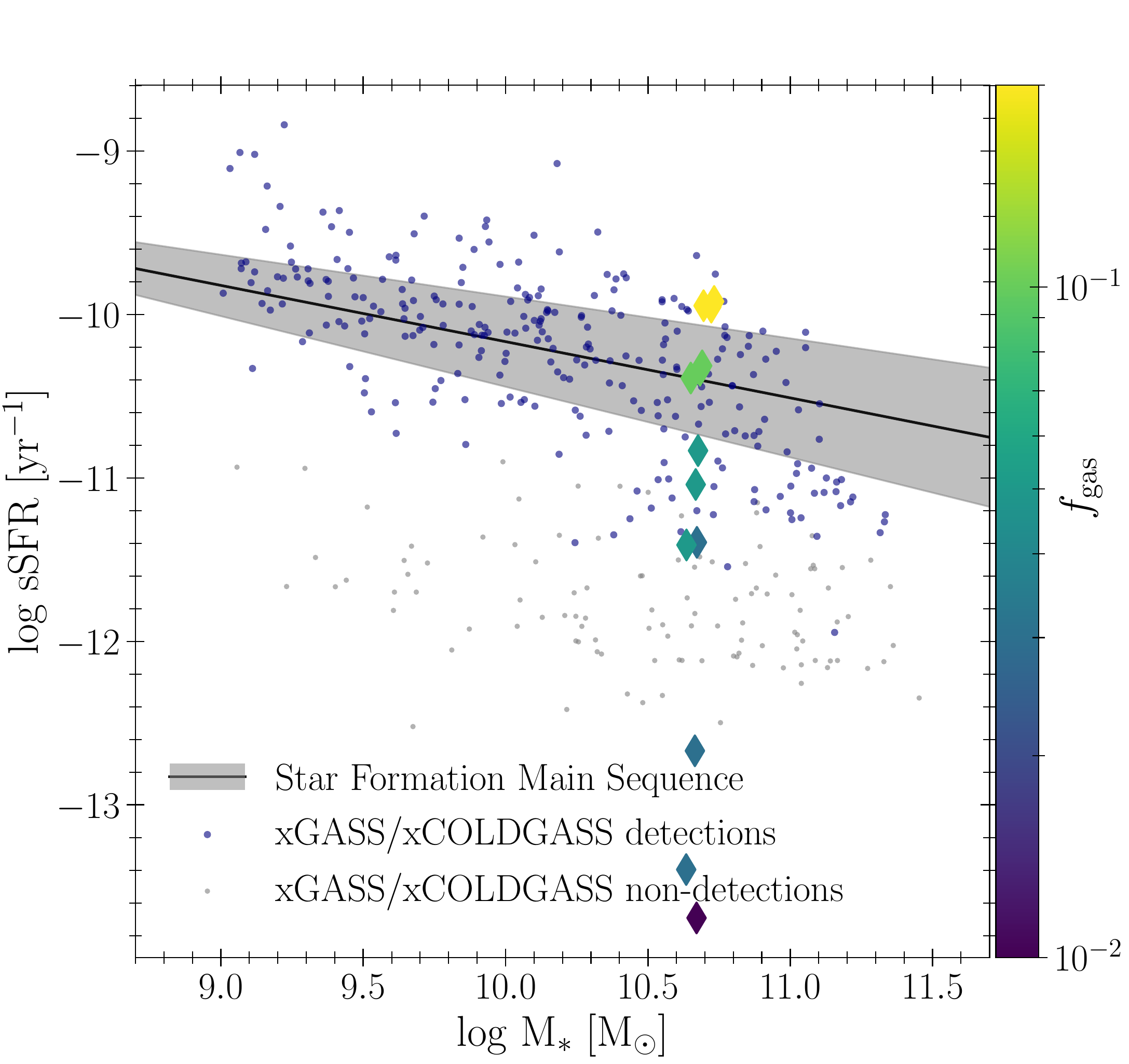}%
    \includegraphics[width=0.47\linewidth]{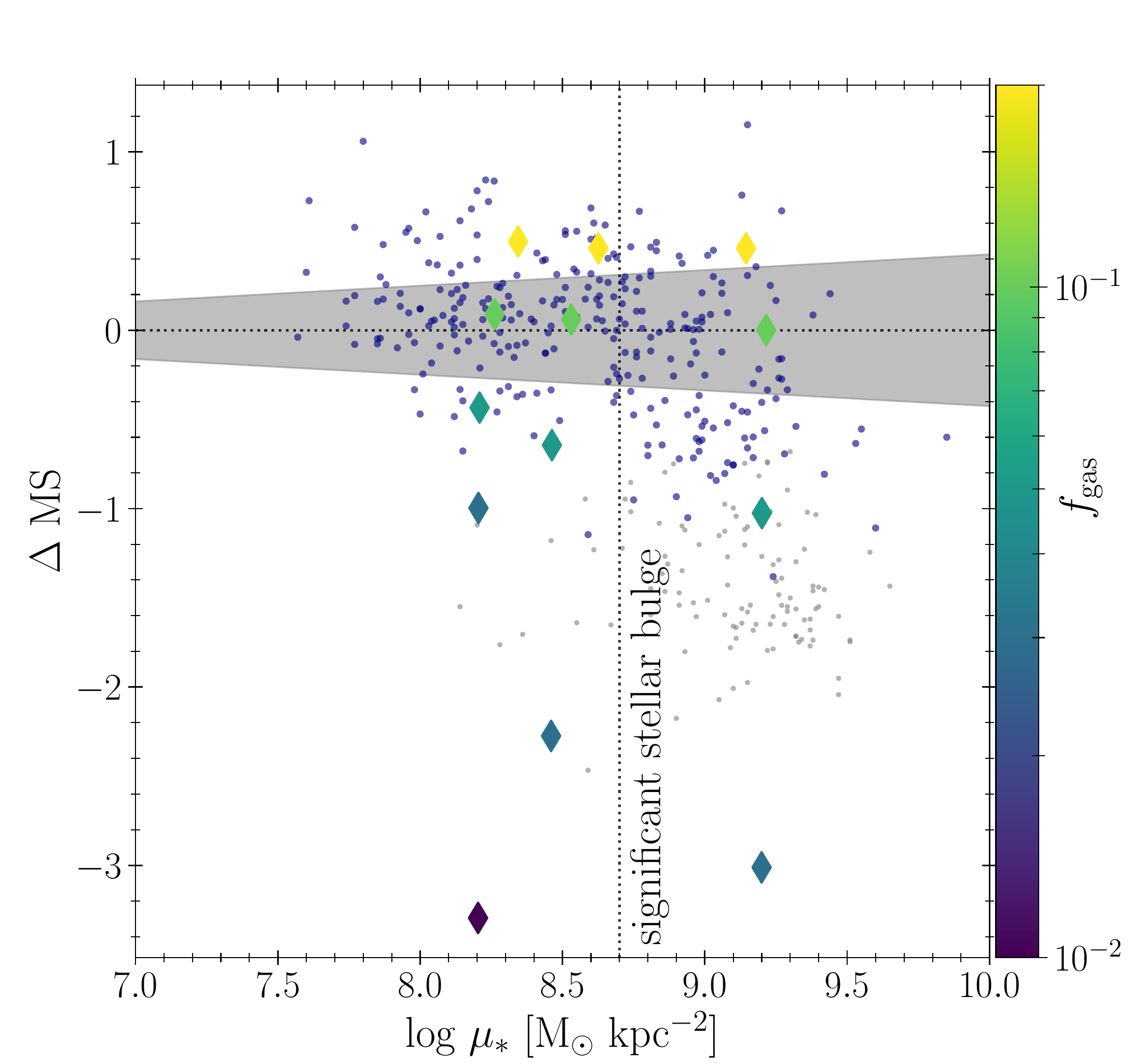}%
    \caption{Comparison of the star formation activity of the simulated galaxies to the observed galaxy population in the xGASS \citet{Catinella2018} and xCOLDGASS \citet{Saintonge2017} surveys. Left: specific SFR (${\rm sSFR}\equiv {\rm SFR}/\Mstar$) as a function of stellar mass. The black line shows the star formation main sequence from \citet{Catinella2018}, with the grey-shaded band indicating the $1\sigma$ scatter. Right: vertical offset from the star formation main sequence as a function of the stellar surface density ($\mus$). In both panels, diamonds show the results from the simulations, colour coded by their gas-to-stellar mass ratio $\fg$. Navy (grey) points show the detections (non-detections) of the xGASS \citep[H{\sc i},][]{Catinella2018} and xCOLDGASS \citep[CO,][]{Saintonge2017} surveys. The central 300$\pc$ of the simulations are excluded from the analysis (see the text). 
    }
    \label{fig:SFMS}
\end{figure*}

\subsection{Star formation} \label{ss:SF}
We now investigate how the integrated star formation activity of a galaxy is affected by the combination of its gas fraction and gravitational potential. In Figure~\ref{fig:SFMS}, the global star formation activity of the simulated galaxies is compared to that of galaxies in the xCOLDGASS \citep{Saintonge2017} and xGASS \citep{Catinella2018} surveys. As in \citet{Gensior2020}, we exclude the central 300 pc of our galaxies from the star formation analysis, because we do not model AGN feedback. This form of feedback would help to disperse gas that accumulates at the centre and reaches very high densities. While the simulations do not include AGN feedback, they also lack the gravitational resolution to accurately model the star formation activity at the very centre of the galaxy. The simulation results represent a global average taken between 300~Myr and 1~Gyr, when the galaxy has settled into equilibrium.

The left panel of Figure~\ref{fig:SFMS} shows the specific SFR (${\rm sSFR}\equiv {\rm SFR}/\Mstar$) as a function of stellar mass, where the colour of the simulation points indicates their initial gas-to-stellar mass ratio $\fg$. All simulated galaxies have a very similar stellar mass by construction; differences arise exclusively due to differences in SFR over the course of the simulation. To first order, the (s)SFR increases with the gas-to-stellar mass ratio, which directly follows from the well-known observation that ${\rm SFR}\propto M_{\rm gas}$ for star-forming main sequence galaxies \citep[e.g.][]{Kennicutt1998,Bigiel2008,Leroy2013}. Quantitatively, galaxies with gas-to-stellar mass ratios $\fg>0.05$ fall within the $1\sigma$ scatter around the xGASS and xCOLDGASS \citet{Catinella2018} star formation main sequence (SFMS, e.g.\ \citealt{Noeske2007, Peng2010}). Galaxies with gas-to-stellar mass ratios $\fg\leq0.05$ tend to fall below the main sequence. While Figure~\ref{fig:SFMS} clearly demonstrates that the gas mass has a strong influence on the global SFR of a galaxy, we are mainly interested in trends with the gravitational potential. The left panel of Figure~\ref{fig:SFMS} shows that the spread in the sSFR of simulations with the same gas fraction increases towards galaxies with lower gas fractions, indicating a stronger impact of dynamical suppression. The most gas-rich simulations ($\fg=0.10{-}0.20$) exhibit a negligible range of $\leq 0.1$~dex in sSFR, irrespective of the gravitational potential. By contrast, galaxies with intermediate gas content ($\fg=0.03{-}0.05$) span a range in sSFR of $\sim$2 and 0.6~dex, respectively, highlighting that the gravitational potential has a major impact on the sSFR (recall that \citealt{Gensior2020} adopted $\fg=0.05$ in all simulations). In the most gas-poor systems ($\fg=0.01$), only the disc-dominated run B\_M30\_R3\_fg1 forms a non-negligible amount of stars.\footnote{To verify that our results are not caused by the absence of gas eligible for star formation in the inner $300{-}700~\pc$ of B\_M90\_R1\_fg1 (see the discussion of Figure~\ref{fig:veldisp_radprof}), we ran an additional simulation with a density threshold of $0.1~\ccm$ and a temperature ceiling of $10^4~\K$ to define gas eligible for star formation. We also ran a simulation with the same star formation criteria as B\_M90\_R1\_fg1, but a factor of three higher resolution. Neither of these additional simulations exhibit any significant differences in sSFR relative to B\_M90\_R1\_fg1.} 

To further quantify how the gravitational potential affects the position of a galaxy relative to the SFMS and to show how this position depends on the gas fraction, the right panel of Figure~\ref{fig:SFMS} shows the offset from the \citet{Catinella2018} SFMS as a function of the stellar surface density $\mus$ (defined as the mean surface density within the galaxy's stellar half-mass radius) for all simulations. The stellar half-mass surface density traces both the bulge surface density and bulge-dominance. This panel clearly visualises the key trends that govern the SFMS offsets visible in the left panel. Galaxies with higher stellar surface densities generally fall further below the SFMS, but this trend becomes shallower at higher gas-to-stellar mass ratios, until it eventually vanishes altogether for $\fg=0.20$. In such gas-rich galaxies, the gas achieves densities high enough that it is self-gravitating and forms stars on a free-fall time-scale. For $\fg=0.10$, the simulations show a very slight decrease in star formation activity towards more bulge-dominated potentials, but the overall decrease is only $\sim0.1$~dex. For lower gas fractions, the dynamical suppression of star formation going from disc-dominated to bulge-dominated potentials is much larger, and continues to steepen as the gas fraction decreases. This is a direct consequence of the behaviour seen in Figure~\ref{fig:veldisp_radprof} and described in Section~\ref{ss:GD}, where we find that only galaxies with $\fg<0.05$ are sensitive to the shear induced by bulge-dominated potentials, and that this sensitivity increases towards lower gas fractions.

These results imply that a dominant spheroidal component in itself is not sufficient to quench a galaxy or maintain quiescence -- only galaxies with gas-to-stellar mass ratios $\fg\leq0.05$ are found to show a significant relation between SFMS offset and stellar surface density. Dynamical suppression is rendered ineffective altogether in more gas-rich galaxies (as applies to the high-redshift galaxy population, to which gas-regulator models have been calibrated). In the next section, we derive analytical expressions for the stellar masses and redshifts at which dynamical suppression becomes important.

\section{When the Elephant steps into the Bathtub: Predicting the onset of dynamical suppression} \label{s:EiB}
Figure~\ref{fig:SFMS} highlights that the effectiveness of dynamical suppression exhibits a strong dependence on the gas fraction and a moderate (but important) dependence on the stellar surface density. We quantify the effect on the sSFR by performing a multi-linear regression\footnote{We discuss the validity of using this parameterisation in Appendix~\ref{A:mlr}.} of the simulated data. This yields the following relation:
\begin{eqnarray}    \label{eq:simsSFR}
    \log (\rm sSFR_{\rm sim} [\Gyr^{-1}])& =  - 1.79 + 4.00\log\left(\frac{\fg}{0.05}\right) \\
    &  - 0.99\log\left(\frac{\mus}{10^8~\Msun~\kpc^{-2}}\right) , \nonumber
\end{eqnarray}
for $8.0<\log{(\mus/\Msun~\kpc^{-2})}<9.5$ and $0.01\leq\fg\leq0.20$. By combining this fit with observed scaling relations describing the galaxy population, we can predict a minimum stellar mass as a function of redshift at which dynamical suppression will become effective. 

First, we rewrite the right-hand side of equation~\ref{eq:simsSFR} in terms of $\log\Mstar$ and $z$. The stellar surface density is given by $\mus=\Mstar/(2\pi R_{\rm e})$ and can be converted to solely depend on stellar mass and redshift by assuming a galaxy mass-radius-redshift relation. For this, we use:
\begin{equation}
    R_{\rm e} = 8.9\kpc (1+z)^{-0.75}(\Mstar/5\times 10^{10}\Msun)^{0.22} ,
    \label{eq:re}
\end{equation}
as obtained for late-type galaxies by \citet{vdWel2014}. Early-type galaxies are found to have smaller radii and larger stellar surface densities, which means that the mass-radius relation for late-type galaxies can indeed be used to derive the limiting case that we are interested in here. Secondly, we use the observed scaling of the gas fraction as a function of redshift and stellar mass \citep{Tacconi2018} to eliminate $\fg$ from equation~(\ref{eq:simsSFR}). This adds further justification to our choice of the late-type mass-radius relation, because the \citet{Tacconi2018} relation was derived for star-forming galaxies. We chose their best fit for a \citet{Speagle2014} SFMS, which gives:
\begin{eqnarray}
    &\log \left( \frac{\fg}{0.05}\right) = 0.11 +  2.49\log(1+z) + 0.52\dMS \\
    & - 0.36\log\left(\frac{\Mstar}{5\times10^{10}~\Msun}\right), \nonumber
   \label{eq:Tac_fgas} 
\end{eqnarray}
where $\dMS$ is the main sequence offset in dex.

We define `effective dynamical suppression' as causing a SFMS offset of $\dMS = -0.5$~dex, which is expressed as a negative to reflect a suppression of star formation. The magnitude of the offset, 0.5~dex, is a compromise between the mass-dependent scatter about the main sequence and the maximum variation of 0.6 dex experienced by galaxies throughout their life on the SFMS \citep{Catinella2018, Tacconi2018}. As such, this definition means that galaxies need to be at least $1\sigma$ below the main sequence for dynamical suppression to be `important'. We keep $\dMS$ constant, because \citet{Speagle2014} find that the scatter around the SFMS varies little across cosmic time. Now, we can evaluate equation~(\ref{eq:simsSFR}) at $\dMS = -0.5$~dex and write:
\begin{eqnarray}
    \log (\rm sSFR_{\rm sim} [\Gyr^{-1}])& =  - 2.39 + 8.48\log(1+z) \\
    &  - 1.99\log\left(\frac{\Mstar}{5 \times 10^{10}~\Msun}\right) . \nonumber
    \label{eq:simsSFRf}
\end{eqnarray}

Requiring $\dMS = - 0.5$ dex as our minimum SFMS offset for equation~\ref{eq:simsSFRf} then allows us to define the condition: 
\begin{equation}
    - 0.5 > \log (\rm sSFR_{\rm sim} [\Gyr^{-1}]) - \log (\rm sSFR_{\rm MS, obs} [\Gyr^{-1}]),
    \label{eq:sim_cond}    
\end{equation}
where we use the \citet{Speagle2014} SFMS to define $\log (\rm sSFR_{\rm MS, obs} [\Gyr^{-1}])$ for self-consistency with the preceding derivation, specifically with the $\fg$ relation from \citet{Tacconi2018}. Therefore, equation~\ref{eq:sim_cond} becomes: 
\begin{eqnarray}
    \label{eq:sim_condf}
    -0.5 >& - 2.39 + 8.48\log(1+z)  - 1.99\log\left(\frac{\Mstar}{5 \times 10^{10}~\Msun}\right)  \\ 
    & - (-0.16-0.026\tc)\times(\log\left(\frac{\Mstar}{5 \times 10^{10}~\Msun}\right)+10.73) \nonumber \\
    & + (6.51 - 0.11\tc) - 9 , \nonumber
\end{eqnarray}
for $9.0<\log{(\Mstar/\Msun)}<11.7$, $8.0<\log{(\mus/\Msun~\kpc^{-2})}<9.5$, and $0.01\leq\fg\leq0.20$. If this condition is satisfied, dynamical suppression is predicted to drive galaxies off the SFMS. The $\tc$ in this equation is the age of the universe in Gyr and can be written as a function of redshift \citep{Tacconi2018}:
\begin{eqnarray}
    &\log\tc = 1.143 - 1.026\log(1+z)  - 0.599\log^2(1+z) \\
    & + 0.528\log^3(1+z) \nonumber,
    \label{eq:tc}
\end{eqnarray}
which assumes a flat $\Lambda$CDM universe with $\Omega_m = 0.3$ and $\rm H_0 = 70~\kms~\rm Mpc^{-1}$.

Finally, we can reorder the condition of equation~\ref{eq:sim_condf} to predict the minimum required stellar mass at a given redshift above which dynamical suppression becomes important: 
\begin{equation}
    \log\left(\frac{\Mstar}{5 \times 10^{10}~\Msun}\right) > \frac{-2.66+8.48\log(1+z)+0.17\tc}{1.83-0.026\tc} ,
    \label{eq:m_min}
\end{equation}
where $\tc$ is defined as a function of redshift according to equation~(\ref{eq:tc}). We find that the condition of equation~\ref{eq:m_min} can be approximated to within $<1$~per~cent by a third-order polynomial fit, providing the minimum stellar mass as a function of redshift: 
\begin{equation}
    \log\left(\frac{\Mstar}{5 \times 10^{10}~\Msun}\right) > -0.21+0.87z-0.11z^2+0.0082z^3 .
    \label{eq:m_min_poly}
\end{equation}
Expressed as a maximum redshift as a function of stellar mass, this equivalently becomes:
\begin{eqnarray}     \label{eq:z_max}
    &z < 0.23 + 1.22\log\left(\frac{\Mstar}{5 \times 10^{10}~\Msun}\right) \\
    &+0.17\log^2\left(\frac{\Mstar}{5 \times 10^{10}~\Msun}\right)+ 0.069\log^3\left(\frac{\Mstar}{5 \times 10^{10}~\Msun}\right)  . \nonumber
\end{eqnarray}

\begin{figure}
    \centering
    \includegraphics[width=1.\linewidth]{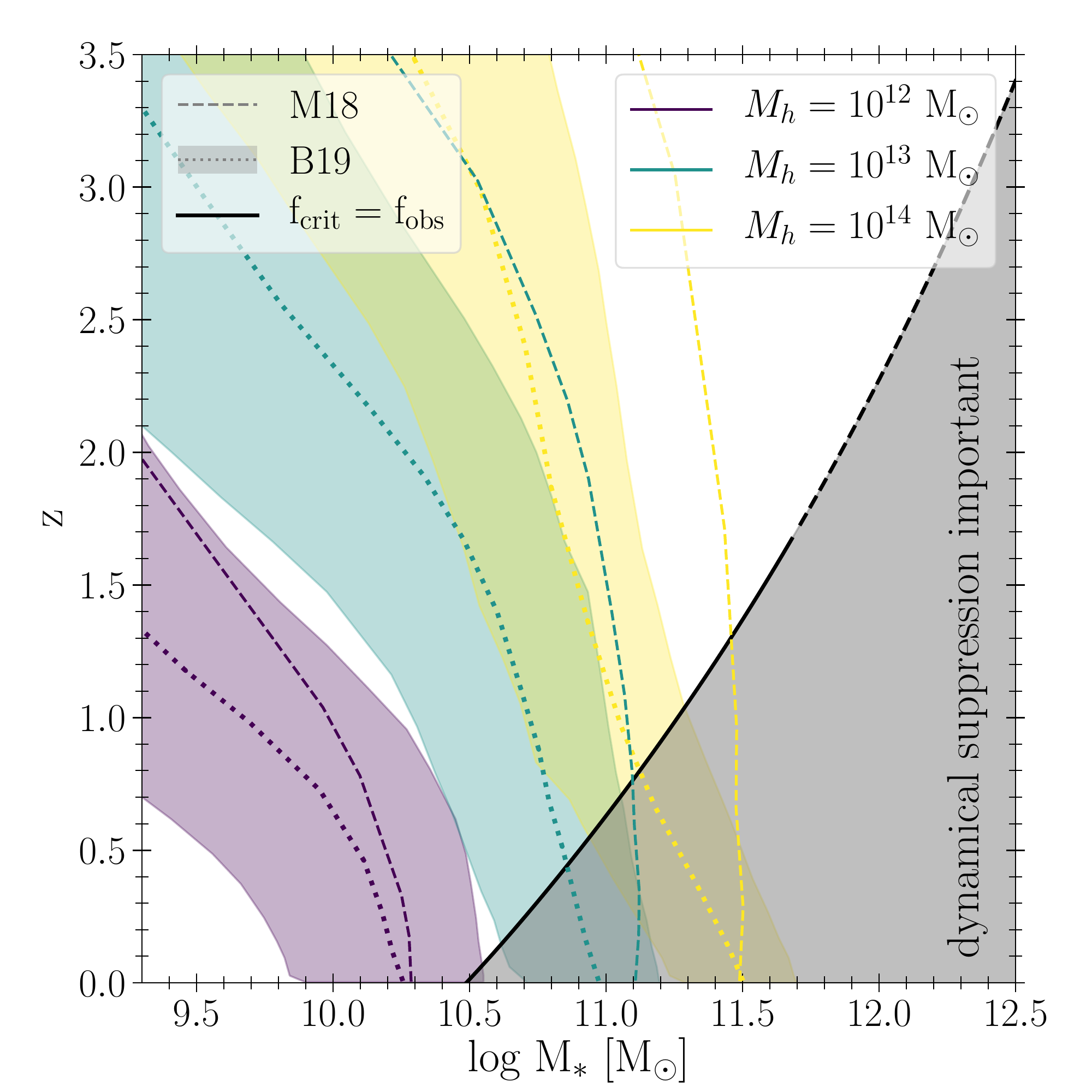}%
    \caption{Redshift and stellar mass range for which galaxies are predicted to be affected by the dynamical suppression of star formation. The black line indicates where the typical observed gas fraction \citep{Tacconi2018} and the gas fraction required for dynamical suppression are equal, as expressed in equation~(\ref{eq:m_min}). Galaxies below the line (grey-shaded area) are predicted to experience dynamical suppression. Coloured lines show the stellar mass growth histories of galaxies with $z = 0$ stellar masses ranging from $1.8\times10^{10}~\Msun$ to $3\times 10^{11}~\Msun$, as predicted by \citet{Moster2018} and \citet{Behroozi2019} (see the legend).
    }
    \label{fig:zmstar}
\end{figure}

The prediction expressed equivalently in equations~(\ref{eq:m_min})--(\ref{eq:z_max}) is shown in Figure~\ref{fig:zmstar}. The black line indicates the minimum stellar mass a galaxy must have at a given redshift (or, conversely, a maximum redshift at a given stellar mass) to experience dynamical suppression, as in equations~(\ref{eq:m_min}) and~(\ref{eq:z_max}). Galaxies residing below the line, in the grey-shaded area, should typically fall below the SFMS as a result of dynamical suppression, even if it does not quench the galaxy completely. As the equation~(\ref{eq:Tac_fgas}) has been derived for stellar masses $\log(\Mstar/\Msun) \leq 11.7$ \citep{Tacconi2018}, the black line is an extrapolation at stellar masses above this limit (visualised by a dashed line).

In Figure~\ref{fig:zmstar}, we show which galaxies are affected by dynamical suppression, and for how long this has been the case. To make this comparison, we adopt the stellar mass assembly histories from \citet{Moster2018} and \citet{Behroozi2019}, who have used a combination of halo abundance matching and empirical constraints. Coloured lines in Figure~\ref{fig:zmstar} show the average assembly history, as predicted by the models of \citet{Moster2018} and \citet{Behroozi2019}, for galaxies with a stellar mass at $z=0$ in the range of $1.8\times 10^{10} \leq \Mstar/\Msun \leq 3\times 10^{11}$, derived from halos with final masses of $10^{12}, 10^{13}$, and $10^{14}\Msun$. The figure shows that $L_{\ast}$ galaxies like the Milky Way have typically been experiencing the onset of dynamical suppression since $z\leq0.23$ (or up to 2.8~Gyr). By contrast, the most massive galaxies considered here ($\Mstar\sim3\times10^{11}~\Msun$) are predicted to have experienced dynamically suppressed star formation since $z\leq0.82$ (up to 6.9~Gyr ago) \citep{Behroozi2019} or $z\le1.36$ (up to 8.8~Gyr ago) \citep{Moster2018} depending on the stellar mass assembly history model.

In the Milky Way, there is evidence for the dynamical suppression of star formation in the bulge-dominated central 500~pc \citep{Longmore2013,Kruijssen2014b}. For galaxies out to $z\sim1$, there is indirect evidence supporting our prediction in terms of observed morphological trends. For instance, the transition from discs to spheroids dominating the stellar mass budgets of galaxies occurs at a stellar mass of $10^{10.5}~\Msun$ for the galaxy population at $z\sim0.1$. Moreover, the average bulge-to-total ratio increases with increasing stellar mass \citep{Thanjavur2016}, such that the most massive galaxies tend to be bulge-dominated from $z\sim1.5$ onward \citep{Bundy2006,Bundy2010,Buitrago2013}. We predict that these galaxies are susceptible to the dynamical suppression of star formation. 

\section{Discussion} \label{s:SD}
\subsection{Summary} \label{ss:Sum}
In this paper, we have used a suite of 15 hydrodynamical simulations of isolated galaxies to investigate under which conditions galactic-dynamical processes can suppress star formation. The simulations span a variety of gravitational potentials (most prominently tracing different stellar surface densities or bulge-to-disc ratios) and gas-to-stellar mass ratios (in the range $\fg=0.01{-}0.20$). The simulations adopt a dynamics-dependent sub-grid star formation model, in which more super-virial gas clouds form stars less efficiently. The results are summarised as follows: 

\begin{enumerate}[leftmargin=0.5cm]
    \item The dynamical suppression of star formation proceeds by the increase of the gas velocity dispersion by shear, rendering the gas supervirial. The bulge component of the simulated galaxies drives a central increase of the velocity dispersion most efficiently at high central stellar surface densities (i.e.\ high bulge-to-disc ratios) and low gas fractions. At the lowest gas-to-stellar mass ratio ($\fg=0.01$), even the weakest bulge component considered in this work is sufficient to drive up the central velocity dispersion. At the highest gas-to-stellar mass ratio ($\fg=0.20$), the median velocity dispersion is insensitive to the gravitational potential. 
    
    \item The extent to which the SFR is dynamically suppressed depends on both the gas-to-stellar mass ratio ($\fg$) and the stellar surface density ($\mus$). For a given $\mus$, we identify the critical $\fg$ above which dynamical suppression becomes ineffective, because the stellar spheroidal component no longer dominates the local gravitational potential. Conversely, the effect of dynamical suppression increases towards lower gas fractions, driving galaxy off the star formation main sequence (SFMS), into the population of quenched galaxies.
    
    \item The gas-to-stellar mass ratio (or gas fraction) does not only act as a threshold for dynamical suppression, but also regulates its overall strength and dependence on stellar surface density. The lower the gas fraction, the steeper the anti-correlation between the SFR and stellar surface density -- not only is the global (s)SFR more suppressed at lower gas fractions, but the amount of suppression compared to the SFMS becomes steeper.
    
    \item We perform a multi-linear regression to quantify the relation between sSFR, $\mus$, and $\fg$. By combining this relation with the observed dependence of these quantities on galaxy mass and redshift \citep{Tacconi2018}, we can derive the subset of the galaxy population for which dynamical suppression is predicted to be effective. We find that the physics of star formation can be the rate-limiting step in the baryon cycle at high galaxy masses (equation~\ref{eq:m_min_poly}) and low redshifts (equation~\ref{eq:z_max}; also see Figure~\ref{fig:zmstar}).  
\end{enumerate}

\subsection{Implications for galaxy evolution and quenching}\label{ss:Imp}
Our findings are in good agreement with recent observational results. Similarly to \citet{Tacconi2018}, who relate the main sequence offset to gas fraction and stellar mass across redshift, nearby grand design spirals also exhibit a direct link between SFR, stellar surface density and gas fraction on 500 pc scales \citep{Morselli2020}. Surveys of the galaxy population at $z\sim0$ additionally suggest that the offset from the SFMS is driven by a combination of (molecular) gas fraction and star formation efficiency \citep[e.g.][]{Zhang2019,Piotrowska2020,Brownson2020}. This is consistent with our findings in Section~\ref{ss:SF}, where we predict that the star formation efficiency decreases steeply with $\mus$ (and can thus be considered synonymous) at fixed gas fraction. Observations also reveal that the main sequence offset increases with bulge-to-disc ratio (\citealt{Bluck2019, Cook2020} and references therein), which again agrees qualitatively with our prediction that high stellar surface densities lead to the dynamical suppression of star formation.

Numerically, our results are in agreement with (but considerably extend) previous simulations of similar systems \citep[e.g.][]{Martig2013,Semenov2016,Su2019,Kretschmer2020}. Like ours, these simulations predict that gas dynamics influence star formation only in sufficiently gas-poor galaxies. Interestingly, \citet{Martig2009} find that star formation in their cosmological zoom-in simulation is dynamically suppressed from $z \sim 0.8$ onward when the galaxy's stellar mass exceeds $10^{11}~\Msun$, which is consistent with our prediction. While the quantitative agreement is likely a coincidence, the general qualitative agreement is encouraging and implies that it is crucial to take dynamical suppression into account when modelling the evolution of massive ($\Mstar>3\times10^{10}~\Msun$) galaxies in the intermediate-to-low redshift universe. The analytical representations of our findings can be readily incorporated in gas regulator models, or as sub-grid models in semi-analytic models and hydrodynamic simulations. 

While the suite of simulations presented here provides key insights into the dynamical suppression of star formation and its dependence on the host galaxy properties, they represent idealised numerical experiments that generate various follow-up questions. The simulations do not model gas inflow onto the galactic halo, nor do the initial conditions include a hot circumgalactic medium. How do more realistic boundary conditions affect the onset and extent of dynamical suppression? The models also do not include AGN feedback, which is a key agent for driving gas outflows and quenching galaxies. What is the interplay between AGN feedback and dynamical suppression? The SN feedback used in the simulations does expel some amount of gas from the disc, which later rains back down onto the galaxy. This provides some level of gas outflow and inflow, despite which dynamical suppression is found to regulate the SFR in an important part of galaxy mass-redshift space. There is reason to believe that the \citet{Padoan2012,Padoan2017} model may somewhat overpredict the extent to which star formation is suppressed \citep{Schruba2019}, because the star formation efficiency carries a strong exponential dependence on the virial parameter. Further uncertainty on the predictions of Figure~\ref{fig:zmstar} may arise through the relations adopted to express the dependence of $\mus$ and $\fg$ on $\Mstar$, particularly due to using an observational $\fg$ relation that solely describes the molecular gas. While the cold ISM in high-redshift galaxies is mostly molecular, there is a transition at $z=0.1{-}0.3$ below which atomic gas becomes the dominant component \citep{Tacconi2018}. This implies that we might overpredict the magnitude and onset of dynamical suppression at galaxy stellar masses $\log (\Mstar/\Msun) \leq 10.75$ and redshifts $z\la0.3$. Nonetheless, this is a relatively small part of the mass range considered in Figure~\ref{fig:zmstar}. The qualitative agreement of our simulations with both observational and numerical studies demonstrates that there is indeed a regime where dynamical suppression drives down the SFR in galaxies. 

Our results do not imply that dynamical suppression is solely responsible for galaxy quenching, even in the grey-shaded area of galaxy mass-redshift space in Figure~\ref{fig:zmstar} where it is expected to act. On its own, the effectiveness of dynamical suppression has a simple dependence on galactic properties -- the higher the gas fraction (or redshift) of a galaxy, the higher its stellar mass must be to have a gravitational potential deep enough to affect the gas motions and star formation. Therefore, dynamical suppression should act more effectively when combined with physical processes capable of preventing the accumulation of a massive gas reservoir. An example of such a process is halo quenching \citep{Dekel2006, Dave2017}, which halts the gas inflow from the cosmic web.\footnote{While halo quenching curtails the growth of the cold ISM from external accretion, the ISM of a galaxy will still be replenished through stellar mass loss from AGB stars. However, recent studies of the cold gas and dust in early-type galaxies show that stellar mass loss only contributes a small fraction to their cold ISM \citep[e.g][]{Davis2019, Kokusho2019}. Therefore, the effectiveness of dynamical suppression will not be strongly affected by the mass loss of old stars.} Alternatively, the existence of a massive gas reservoir could be prevented by AGN feedback, by heating and expelling gas from the galaxy \citep[e.g.][]{Croton2006,Bower2006,Fabian2012,Donnari2019,Lacerda2020,Su2020}. AGN feedback and dynamical suppression might be forming a symbiotic cycle -- dynamical suppression requires some mechanism to deplete the galactic gas supply (a role plausibly fulfilled by AGN feedback), and once this is achieved, dynamical suppression will help preserve and build a quiescent gas disc at the centre of the galaxy, which is not consumed by star formation and therefore can feed the AGN again. Our simulations suggest that this nuanced picture, in which a combination of dynamical suppression and AGN feedback or halo quenching regulates the baryon cycle, is necessary to create the quiescent galaxy population \citep[as been argued by a number of observational studies, see e.g.][]{Lang2014,Hahn2017}. 

Ultimately, our prediction that the dynamical suppression of star formation should manifest itself at certain galaxy masses and redshifts (Figure~\ref{fig:zmstar}) should be interpreted as evidence for a regime in which galaxy evolution is no longer solely determined by the balance between gas inflow and outflow. Recently, there have been attempts to add improved star formation physics to gas-regulator models. For instance, \citet{Belfiore2019} include an empirically-motivated star formation efficiency that changes as a function of galactic radius, whereas \citet{Tacchella2020} take into account molecular cloud lifetimes. Figure~\ref{fig:zmstar} acts as a guide to determine on which parts of parameter space future studies should focus, both in observational surveys of the galaxy population and through cosmological (zoom-in) simulations. 

\section*{Acknowledgements}
We thank Volker Springel for allowing us access to {\sc{Arepo}} and Frederic Bournaud for his helpful and constructive referee report. JG thanks Timothy Davis and the members of the MUSTANG group, particularly Benjamin Keller, for helpful discussions.
The authors acknowledge support by the High Performance and Cloud Computing Group at the Zentrum f{\"u}r Datenverarbeitung of the University of T{\"u}bingen, the state of Baden-W{\"u}rttemberg through bwHPC and the German Research Foundation (DFG) through grant no INST 37/935-1 FUGG. JG and JMDK gratefully acknowledge funding from the Deutsche Forschungsgemeinschaft (DFG, German Research Foundation) through an Emmy Noether Research Group (grant number KR4801/1-1). JMDK gratefully acknowledges funding from the DFG Sachbeihilfe (grant number KR4801/2-1). JMDK gratefully acknowledges funding from the European Research Council (ERC) under the European Union's Horizon 2020 research and innovation programme via the ERC Starting Grant MUSTANG (grant agreement number 714907).

\section*{Data availability}
The data underlying this article will be shared on reasonable request to the corresponding author.





\bibliographystyle{mnras}
\bibliography{references}



\appendix

\section{Validation of the multi-linear regression} \label{A:mlr}

\begin{figure}
    \centering
    \includegraphics[width=1.\linewidth]{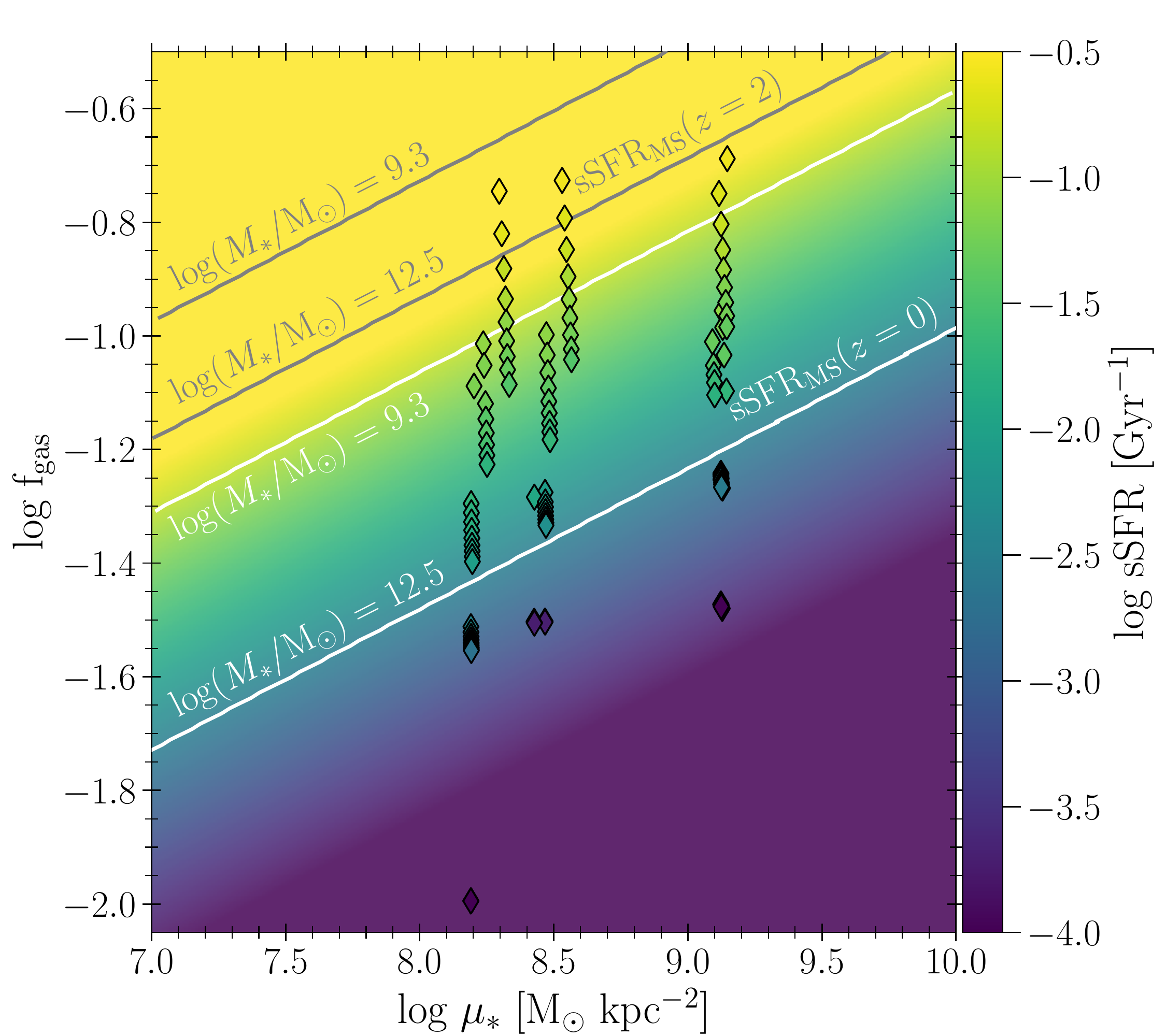}%
    \caption{Gas-to-stellar mass ratio $\fg$ and stellar surface density ($\mus$) plane, coloured by the specific SFR (${\rm sSFR}\equiv {\rm SFR}/\Mstar$) predicted by the multi-linear regression of equation~(\ref{eq:simsSFR}). Diamonds show results from the simulations at different times between 0.3--1~Gyr, colour-coded by their current sSFR. The grey and white lines indicate the expected range in sSFR between stellar masses $2\times10^9 \Msun \leq \Mstar \leq 3\times10^{12}\Msun$ for galaxies on the \citet{Speagle2014} SFMS at redshift 2 and 0 respectively. The good agreement between the colours of the data points and the background demonstrates that the multi-linear regression provides a satisfactory description of the simulations.
    }
    \label{fig:mlr}
\end{figure}
Figure~\ref{fig:mlr} compares the sSFR measured in the simulations to the prediction from the multi-linear regression of equation~(\ref{eq:simsSFR}), for a range of gas-to-stellar mass ratios and stellar surface densities. The comparison between the colour of the diamonds and the background shading shows good agreement between the two, within the boundaries set by the grey and white lines, which demarcate the range in sSFR expected from galaxies on the \citet{Speagle2014} SFMS with stellar masses $2\times10^9 \Msun \leq \Mstar \leq 3\times10^{12}\Msun$. We require agreement between the simulation data-points and the fit in the part of $\fg {-} \mus$ parameter space encased by the lines, because this is the mass range considered in our prediction for the importance of dynamical suppression on the galaxy population. The grey lines denote values for the main sequence sSFR at $z=2$, while the white lines show the sSFR for $z=0$ galaxies. Consequently, the assessment of the quality of the fit should be restricted to points above the $\log\Mstar = 3  \times 10^{12}\Msun, z=0$ white line. Figure~\ref{fig:mlr} shows excellent agreement between linear model and simulations in this part of parameter space. Across the entire model grid, only a few points show a lower sSFR compared to the multi-linear regression fit, likely because our sub-grid star formation model might overestimate the suppression of star formation somewhat in part of parameter space (see discussion in Section~\ref{ss:Imp}). However, these points fall below the bottom white line, i.e.\ outside of our region of relevance, and thus do not influence our conclusion that a multi-linear parametrisation of the sSFR is an appropriate fit to the simulated data.

\section{Gas density distributions} \label{A:gas}
\begin{figure*}
    \centering
    \includegraphics[width=1.\linewidth]{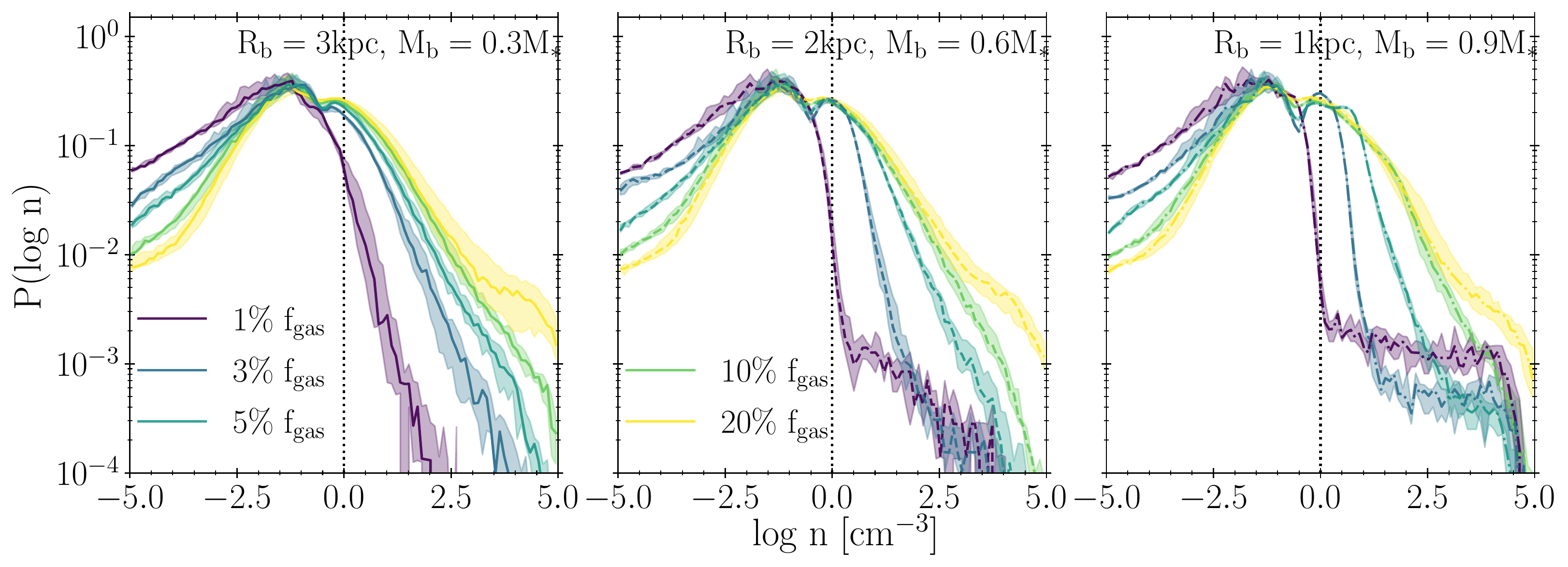}%
    \caption{Gas volume density distributions of the simulated galaxies. Each panel shows the probability density function (PDF) of the gas for galaxies with gas-to-stellar mass ratios of $0.01{-}0.20$, for the disc-dominated (left), intermediate bulge (middle) and bulge-dominated (right) potentials. Lines indicate the median over time, while the shaded regions indicate the 16th-to-84th percentile variation over time. The vertical black dotted line indicates the minimum density threshold for star formation, which is $n_{\rm min} = 1~\ccm$.
    }
    \label{fig:pdfs}
\end{figure*}

\begin{figure*}
    \centering
    \includegraphics[width=0.82\linewidth]{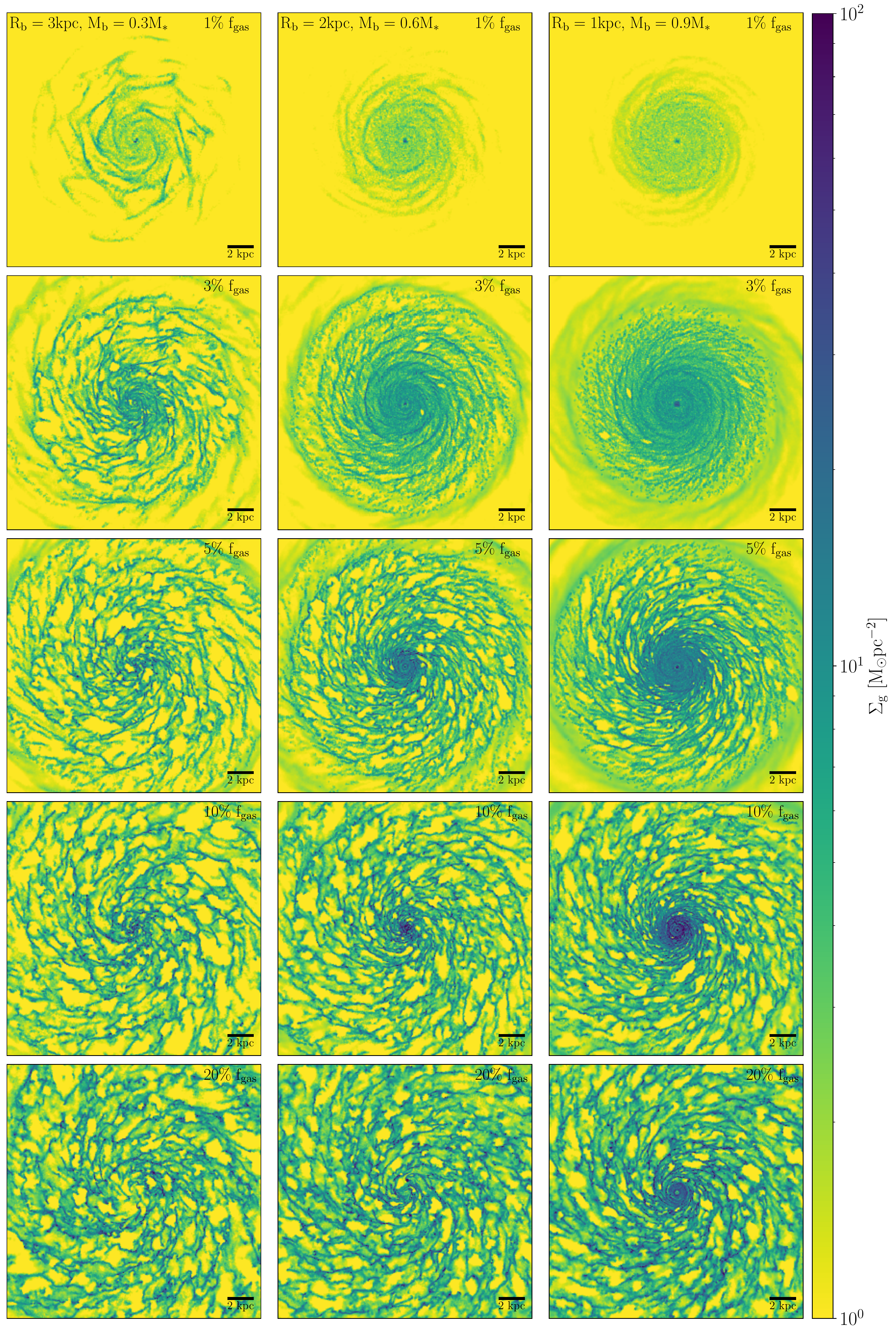}%
    \caption{Surface density projection of the gas component in the simulated galaxies. The columns indicate the distinct gravitational potentials, with disc-dominated on the left, intermediate bulge in the middle and bulge-dominated on the right. The gas-to-stellar mass ratio increases from 1~per~cent in the top row to 20~per~cent in the bottom row. The maps are shown at 900~Myr after the start of the each simulation and measure $20~\kpc$ on a side. At low gas-to-stellar mass ratios, we find ubiquitous smooth nuclear gas discs, which arise from the dynamical suppression of fragmentation. At $\fg \geq 10$~per~cent, the ISM is predominantly substructured, due to gravitational instability and stellar feedback. At such a high gas fraction, even the completely bulge-dominated potential only hosts a very small central disc.
    }
    \label{fig:SDmaps}
\end{figure*}

The probability density function (PDF) for all gas within the simulations is shown in Figure~\ref{fig:pdfs}, for the complete set of different stellar density profiles and gas-to-stellar mass ratio. The main difference in the gas distribution between the different gas fractions lies in the high-density tail of the PDF. The higher the gas-to-stellar mass ratio, the more gas is present at high densities. This difference between galaxies with different gas-to-stellar mass fractions increases the more spheroid-dominated the gravitational potential is. It is caused by shear, which inhibits the fragmentation of gas into smaller, denser clouds at low gas-to-stellar mass ratios (see also Figure~\ref{fig:SDmaps}). Despite these differences at low gas-to-stellar mass ratios, the mass-weighted fractions of gas above the star formation threshold for the runs with $\fg=0.05{-}0.20$ only differ by a couple of per cent or less. The sharp drop of the gas PDF just below the density threshold for star formation seen for runs B\_M60\_R2\_fg1 and B\_M90\_R1\_fg1 might suggest a numerical bias in the sSFR of these galaxies, related to the choice of threshold density. To address this, we performed two (re-)simulations of B\_M90\_R1\_fg1 (one with a density threshold of $n_{\rm min} = 0.1~\ccm$ and one with higher resolution), which show that the SFRs presented in this paper are not affected by the threshold choice and resolution.

Figure~\ref{fig:SDmaps} shows a projection of the gas surface density for each simulated galaxy. \citet{Gensior2020} found that the shear induced by a dominant spheroidal component can suppress fragmentation in the ISM of the galaxy, creating a smooth circumnuclear gas disc. The extent of the disc depends on the bulge strength, i.e.\ the higher the central stellar surface density, the larger disc. Figure~\ref{fig:SDmaps} demonstrates that with $\fg \leq 3$~per~cent even the disc-dominated galaxies show signs of this process, even if the central gas reservoir is marginally disturbed. The more gas dominates the local gravitational potential at higher gas-to-stellar mass ratios, the lower the impact of the stellar potential. As shear becomes insufficient in suppressing fragmentation, star formation and stellar feedback proceed. These processes together lead to an increasingly more substructured ISM. At a gas-to-stellar mass ratio of 20~per~cent, the most bulge-dominated potential barely hosts a central gas disc. Figure~\ref{fig:SDmaps} illustrates that, similarly to the dynamical suppression of star formation, the extent and existence of the smooth central gas disc also depends on both the gas-to-stellar mass ratio and the underlying gravitational potential.


\bsp	
\label{lastpage}
\end{document}